# In-field dependences of the critical current density $J_c$ in GdBa$_2$Cu$_3$O$_{7-d}$ coated conductors produced by Zr irradiation and post-annealing at low temperatures

N. Haberkorn,[1,2] S. Suárez,[1,2] Jae-Hun Lee,[3] S. H. Moon,[3] Hunju Lee.[3]

[1] Comisión Nacional de Energía Atómica and Consejo Nacional de Investigaciones Científicas y Técnicas, Centro Atómico Bariloche, Av. Bustillo 9500, 8400 San Carlos de Bariloche, Argentina.

[2] Instituto Balseiro, Universidad Nacional de Cuyo, Av. Bustillo 9500, 8400 San Carlos de Bariloche, Argentina

[3] SuNAM Co. Ltd, Ansung, Gyeonggi-Do 430-817, South Korea.

*e-mail*: *nhaberk@cab.cnea.gov.ar*  Tel: +54-294-444-5147- FAX: +54-294 444-5299.

*Keywords:* A. Coated conductors; D. Critical Current Densities; E. Ion Irradiation.

We report the influence of 6 MeV Zr$^{4+}$irradiation and post-irradiation annealing (200 °C) in the in-field dependences of the critical current densities $J_c$ of 1.3 thick GdBa$_2$Cu$_3$O$_{7-d}$ coated conductors grown by co-evaporation. Samples were irradiated with 6 MeV Zr$^{4+}$ and fluences between 2.3x10$^{11}$ cm$^{-2}$ and 3x10$^{12}$ cm$^{-2}$. The correlation between the superconducting critical temperature $T_c$ and in-field dependences of critical current densities $J_c$ has been analyzed. In addition, random disorder introduced by irradiation was reduced by thermal annealed at 200 °C. The analysis of our experimental findings indicates that the optimal irradiation (reducing random disorder by annealing) results in the suppression of the self-field $J_c$ of ≈ 10 % and in- field $J_c$ enhancements nearly doubled at about 5 T. A clear correlation between $T_c$, disorder and self-field $J_c$ is observed. Additional random disorder and nanoclusters suppress systematically $T_c$ and increase the flux creep relaxation at intermediate temperatures (reducing the characteristic glassy μ value).

## 1. Introduction

Artificially designed mixed pinning landscapes seem to be a tool to improve critical current densities $J_c$ in high temperature superconductors [1,2,3]. A significant enhancement of the in-field $J_c$ of $R$Ba$_2$Cu$_3$O$_{7-\delta}$ ($R$BCO; $R$: Sm, Dy, Y, Gd) coated superconductors (CCs) can be obtained by adding pinning centers by ion irradiation [4,5,6,7,8]. For adequate irradiation fluences (which depend on mass and energy of ions), adding small clusters and random disorder assists the pinning produced by normal inclusions and twin boundaries (originated during the synthesis) [9,10]. The optimal doses for irradiation result from a balance between the retention of intrinsic superconducting properties and the enhancement of the vortex pinning. The increment of the disorder at the nanoscale reduces systematically the superconducting critical temperature ($T_c$) and the self-field $J_c$. In addition, the decay overtime of the persistent currents (flux creep rates) at intermediate and high temperatures (> 20K) displays higher values when the irradiation fluency is increased [4,5,6]. This change in the vortex dynamics has been related to the influence of mixed pinning landscapes in the vortex bundle size [11]. On the other hand, we have recently demonstrated that the reduction in $T_c$ (by changing the oxygen stoichiometry) also increases the flux creep rates in CCs [12].

In this work we study the influence of 6 MeV Zr$^{+4}$ irradiations in the in-field $J_c$ dependences and the vortex dynamics of 1.3 μm thick GBCO films by performing magnetization measurements. For comparison (removal of random disorder), similar measurements were performed after annealing the films at 200 °C for 30 minutes. The objective of this study is to find a correlation among the disorder produced by irradiation, the $T_c$ and the self-field and in-field dependence of $J_c$. The pristine films display a pinning landscape with sphere-like and irregular precipitates (Gd$_2$O$_3$) embedded in the GBCO matrix with typical diameter of approximately 50 nm [7]. In addition, correlated pinning produced by twin boundaries and boundaries between islands usually assists the pinning produced by the nanoparticles [13]. For proton and oxygen irradiation, the optimal doses result from a balance between the enhancement of the critical current densities $J_c$ and the suppression of the superconducting properties produced by the damage (reduces self-field $J_c$). Negligible contribution of oxygen vacancies (considered random point defects) to the vortex pinning of pristine 1.3 μm thick GBCO coated conductors was previously observed [12]. Assuming that irradiation with energies of a few MeV produces nanoclusters and random point defects (with larger influence on $T_c$) [4,6] a question arises: is it possible to increase $T_c$ and maximize vortex pinning (reducing vortex fluctuations) by reducing random disorder? Following this hypothesis, the $J_c(H)$ dependences (5 K, 27 K and 40 K) were measured in GBCO CCs irradiated with 6 MeV $Zr^{+4}$ and then a post-irradiation annealing (200 °C) was done. In addition, the vortex dynamics was analyzed by performing magnetic flux creep measurements.

## 2. Experimental

The GBCO tape was grown by the co-evaporation technique previously described in ref. [14]. The magnetization (**M**) measurements were performed by using a superconducting quantum interference device (SQUID) magnetometer with the applied magnetic field (**H**) parallel to the *c*-axis (**H**$\parallel$*c*). The $J_c$ values were calculated from the magnetization data using the appropriate geometrical factor in the Bean Model. For **H**$\parallel$c, $J_c = \frac{20\Delta M}{w\left(1 - w/3l\right)}$, where $\Delta M$ is the difference in magnetization between the top and bottom branches of the hysteresis loop, $l$ and $w$ are the length and width of the film ( $l > w$), respectively. The measurements were recorded for more than 1 hour. The initial time was adjusted considering the best correlation factor in the log-log fitting of the $J_c(t)$ dependence. The initial critical state for each measurement was generated using $\Delta H \sim 4H^*$, where $H^*$ is the field for full-flux penetration [15].

Irradiation with 6 MeV $Zr^{+4}$ is expected to produce random point defects and nm-sized anisotropic defects. Previous irradiation studies of GBCO CCs showed that the optimal dosage to achieve the maximum pinning enhancement at temperatures below 40 K, using either H or O ions are $2\times10^{16}$ $cm^{-2}$ and $1\times10^{14}$ $cm^{-2}$, respectively [5,7]. However, according to SRIM simulations, the Zr ions will produce significantly more displacements per collision than H or O. Thus, the required dosage should be significantly less. The irradiation was performed at room temperature on pieces with typical area 1.2 x 1.2 mm using ion beam currents between 0.15 and 2.4 nA. The samples were irradiated with the ion beam oriented along the crystallographic *c*-axis of the GBCO. To guarantee proper thermal contact, the samples were fixed to the holder with silver paint. The irradiation spot was 1.5 mm in diameter. All the samples displayed similar properties before irradiation. Wherever used, the notation IRRx indicates a GBCO film without irradiation (x=0), whereas x = 2.3, 3.5, 7, 17, and 30 correspond to films irradiated with $Zr^{4+}$ fluence of 2.3x10$^{11}$ cm$^{-2}$, 3.5x10$^{11}$ cm$^{-2}$, 7x10$^{11}$ cm$^{-2}$, 1.7x10$^{12}$ cm$^{-2}$ and 3x10$^{12}$ cm$^{-2}$, respectively. After irradiation and for a better comparison of the properties, IRR3.5 andIRR7 were annealed at 200 °C. As no appreciable differences were observed between 30 min and 180 min of annealing, the standard time for this process was established at 30 min. The notation IRRxA corresponds to samples irradiated with doses X and annealed at 200°C for 30 min.

## 3. Results and discussions

Figure 1 shows the evolution of $T_c$ for IRR0, IRR3.5, IRR3.5A, IRR7 and IRR7A as determined from the magnetic transition measured in $\mu_0 H$ = 0.5 mT (applied after zero-field cooling). Straight lines represent irradiated samples (IRR3.5 and IRR7) and dotted lines, samples annealed at 200°C (IRR3.5A and IRR7A). Table 1 shows a summary of the $T_c$ values for all studied films. As expected, $T_c$ is systematically suppressed when the irradiation fluence is increased. The thermal annealing at 200°C slightly increases the $T_c$ and also produces an abrupt magnetic flux penetration at the superconducting transition. This can be associated with a reduction of the random disorder, which increases the vortex pinning due to a reduction in the vortex fluctuations close to $T_c$ (associated to variations in λ ($T \rightarrow T_c$) [16]).

Figure 2*a* shows on log-log scales the field-dependence of the critical current density $J_c$ (from the Bean model) for IRR0, IRR3.5, IRR7 and IRR17. The results show that an increment in the irradiation fluence systematically decreases $J_c(H{\rightarrow}0)$ and produces smooth $J_c(H)$ dependences. With adequate doses (IRR3.5 and IRR7), the in-field dependence of $J_c$ is improved, almost doubled at around 5 T. In addition, the comparison between IRR7 and IRR7A (see inset Fig. 2*a*) shows that thermal annealing increases $J_c(H{\rightarrow}0)$ and produces faster decays of $J_c$ ($H$), which result in similar $J_c$ values at $\mu_0 H$ =5 T. Following, the in-field dependence of $J_c$ (5 K, 27 K and 40 K) at $\mu_0 H$> 0.3 T will be analyzed. For simplicity, $J_c$ ($H$) is approximated as a power-law regime ($J_c{\propto}H^{-\alpha}$). The dose dependences of α, $J_c^{H{\rightarrow}0}$ and $J_c^{5T}$ after Zr irradiation are presented in Fig. 2*c-d*. The results indicate that α decreases with the fluence ($\approx$ 0.4 for IRR7) and similar changes take place at 5 K, 27 K and 40 K. For IRR7, $J_c^{5T}$ is nearly doubled compared to IRR0, whereas $J_c^{H{\rightarrow}0}$ is almost 25 % lower (i.e. from $\approx$ 33 MA cm$^{-2}$ to $\approx$ 26 MA cm$^{-2}$ at 5 K). Zr-irradiation

fluences higher than IRR7 reduce significantly $J_c^{H\to0}$ with negligible effect in the $\alpha$ value. The correlation between $\alpha$ and $J_c$ values at small and high magnetic fields is similar to those found in proton and oxygen irradiation [4-7]. Usually, the inclusion of random disorder and nanoclusters makes worse the $J_c$ values at low fields. At this state the vortices remain pinned to the large defects and the systematic $J_c^{H\to0}$ drop may be attributed to changes in the superfluid density due to an increment in the disorder at the nanoscale [16]. Smooth $J_c(H)$ dependences evidence that for high magnetic fields, adding random disorder and nanoclusters modifies the strong pinning regimes. For random nanoparticles, when interstitial vortices appear, $J_c$ is expected to vary approximately from $H^{1/2}$ to $H^1$ [2,17]. Initially, our samples display a pinning landscape with large precipitates (typically 50 nm) and twin boundaries [6], which result in values $\alpha\approx0.7$. After being irradiated, they systematically drop to $\alpha\approx$ 0.5-0.4. As mentioned in the introduction, we have shown that oxygen vacancies do not contribute to pinning in oxygen deficient CCs [12]. In this context, random disorder at IRR3.5 and IRR7 was reduced by thermal annealing in pure oxygen at 200 °C. For 5 K, 27 K and 40 K (see Fig. 2$c$-$d$), the $J_c^{H\to0}$ values increased (in agreement with an increment in $T_c$), but the $J_c^{5T}$ was similar to those observed before the annealing. This indicates that random disorder contributes mainly to reduce the performance of the CCs at low magnetic fields (where the vortices are mainly pinned by large defects) but has a negligible contribution in magnetic fields above 5T.

Figure 3 shows the temperature dependence of the normalized logarithmic flux creep rate, $S = -\delta\ln J_c/\delta\ln t$, at 0.5 T for various irradiation fluences. The qualitative features of the $S(T)$ curves are similar to previous observations in $R$BCO [15]. The initial increase of $S(T)$ corresponds to an Anderson-Kim like creep with $S\approx T/U$, where $U$ is the activation energy (approximately T-independent at low T). The IRR0 displays a peak at $T\approx20$ K, usually attributed to correlated disorder (such as twin boundaries) [9], but which also can be attributed to changes in the strong pinning regimes for nanoparticles [18]. Above the peak, the $S(T)$ relaxation displays a minimum which is usually attributed to glassy relaxation with values determined by the vortex bundle size. Irradiation systematically suppresses the peak at $T\approx$ 20 K. Furthermore, at low temperatures, it provokes a reduction in the $S$ values whereas it has a contrary effect at intermediate and high temperatures. According to the collective creep theory, the dynamics in a glassy vortex phase [19] is described by an effective activation energy as a function of current density ($J$) $U_{eff} =$ $\frac{U_0(T)}{\mu}\left[\left(\frac{J_c}{J}\right)^\mu - 1\right]$ [1], where $U_0(T) = U_0 G(T)$ is the scale of the pinning energy, $U_0$ is the collective pinning barrier at $T$=0 in the absence of a driving force, $G(T)$ contains the temperature ($T$) dependence of the superconducting parameters, and $\mu > 0$ is the regime-dependent glassy exponent determined by the bundle size and vortex lattice elasticity. From eq. [1], the temperature dependence of the creep rate ($S$) results in $S = -\frac{d(\ln J)}{d(\ln t)} = \frac{T}{U_0 + \mu T \ln(t/t_0)} = \frac{T}{U_0}\left(\frac{J}{J_c}\right)^\mu$ [2], where $t_0$ is a vortex hopping attempt time. This equation predicts that, with increasing temperature, the second term in the denominator dominates $U_0$, and S approaches the limit $S \approx \frac{1}{\mu \ln \frac{t}{t_0}}$ [15]. Based on the model of nucleation of vortex loops, for random point defects in the three-dimensional case, $\mu$ is 1/7, 3/2 or 5/2, and 7/9 for single vortex, small-bundle and large-bundle creep, respectively [19]. The effective activation energy $U_{eff}(J)$ can be experimentally obtained considering the approximation in which the current density decays as $\frac{dJ}{dt} = -\left(\frac{J_c}{\tau}\right)e^{-\frac{U_{eff}(J)}{T}}$. The final equation for the pinning energy is $U_{eff} = -T\left[\ln\left|\frac{dJ}{dt}\right| - C\right]$ (with C, a constant factor)[20]. For an overall analysis it is necessary to consider $G(T)$, which results in $U_{eff}(J, 0) \approx U_{eff}(J, T)/G(T)$ [21]. Figure 3$b$ shows the Maley analyses for IRR7 and IRR7A. Similar analyses were performed for all studied samples. The inset corresponds to the used $G$ $(T)$ dependence. In the limit of $J << J_c$ the $\mu$ exponent can be estimated as $\Delta$ln $U(J)$ / $\Delta$ln $J$ [22]. The $\mu$ values obtained at intermediate temperatures are summarized in Table I. Initially, IRR0 displayed a $\mu$ = 1.63 that systematically decreased to 1.28 for IRR17. The $\mu$ value observed in the pristine sample is similar to those observed in other CCs [23]. The IRR7 and IRR7A display $\mu \approx$ 1.4-1.5, which are in agreement with previous results for proton and oxygen irradiation. The thermal annealing produces a slight increment in $\mu$ with values between irradiated and pristine samples.

Finally, the influence of the flux creep rates in the single vortex regime (SVR) will be analyzed. SRV refers to negligible vortex-vortex interaction compared to vortex-defect interaction [19]. The pinning in type II superconductors may originate in disorder in $T_c$ ($\delta T_c$) and/or from the spatial variation in the free path $l$ near a lattice defect ($\delta l$) [24]. In the SVR, $J_c$ can be expressed as function of the temperature as $J_c \propto (1 - \left(\dfrac{T}{T_c}\right)^2)^n$, where the exponent $n$ indicates the type of pinning, being 7/6 and 5/2 for $\delta T_c$ and $\delta l$ pinning, respectively [25]. Intermediate values have been observed for mixed pinning landscapes (depending on nanoparticle size and density). For example, in YGdBa$_2$Cu$_3$O$_y$ films grown by metal organic deposition, $n = 1.24$ and 1.55 for nanoparticles with diameters of $\approx 20$ nm and $\approx 90$ nm, respectively. Figure 4 shows $J_c$ vs. $(1 - \left(\dfrac{T}{T_c}\right)^2)$ at $\mu_0 H = 0.1$ T, the data fit with $n = 1.55$ for IRR0, and systematic shift to $n = 1.76$ for IRR7. The $n = 1.55$ value approximates to what would be theoretically expected for $\delta T_c$ pinning, and the systematic increment in $n$ seems indicative of larger contribution of random disorder to the pinning ($\delta l$). This fact can be related to the pinning for the vortex segments between large nanoparticles, which interacts mainly with small defects produced by the irradiation and produces poorer retention in $J_c$ ($T$) (possibly associated to larger thermal fluctuations by local changes in the penetration depth $\lambda$).

## 4. Summary

We examined the influence of 6 MeV Zr$^{+4}$ irradiation effect in a 1.3 μm thick GBCO coated conductors grown by co-evaporation. The results show that the optimal fluence to enhance the in-field dependence is around $7.0 \times 10^{11}$ cm$^{-2}$. For this fluence and in comparison with the pristine film, the $J_c^{sf}$ is reduced to around 25 % (from $\approx 33$ MA cm$^{-2}$ to $\approx 26$ MA cm$^{-2}$ at 5 K), whereas its value is nearly doubled at about 5 T. Thermal annealing at 200 °C reduces the random disorder and the $T_c$ is increased. After annealing and at 5K, $J_c^{sf}$ is increased to $\approx 30$ MA.cm$^{-2}$, which indicates a large contribution of random disorder to the suppression of the $J_c$ values at low fields. However, the reduction of random disorder produces poorer in-field dependences, which results in similar $J_c$ values at 5 T. For all the analyzed irradiated fluences, larger contribution of random disorder and increment of pinning associated with fluctuations in the mean free path ($\delta l$) were observed. The optimal irradiation (considering reduction of random disorder by annealing) results in the suppression of the $J_c^{sf}$ of $\approx 10$ % and $J_c$ enhancements almost doubled at about 5 T.

A clear correlation between $T_c$, disorder and $J_c^{sf}$ is observed. In addition, at intermediate temperatures, the flux creep relaxation is systematically affected by the suppression in $T_c$ by reducing the characteristic glassy µ value.

## Acknowledgement

This work has been partially supported by ANCYPT PICT 2015–2171.

Table I. Summary of Zr irradiation fluences, superconducting critical temperature ($T_c$) and glassy exponents.

Figure 1. Temperature dependence of the magnetization (µ$_0H$ = 0.5 mT) in the pristine and the irradiated films. The data for the irradiated films after be annealed 200 °C was included. The magnetization value was normalized by its value at 60 K.

Figure 2. *a)* Magnetic field dependence of the $J_c$ for IRR0, IRR3.5, IRR7 and IRR17 at 5 K; *b)* α versus Zr fluence obtained from $J_c(H) \propto H^{-\alpha}$ at µ$_0H$> 0.3 T; *c-d)* Self-field $J_c$ and $J_c^{5\,T}$ versus Zr fluence (for 5 K, 27 K and 40 K), respectively.

Figure 3. *a)* Temperature dependence of the creep flux rate ($S$) at µ$_0$H = 0.5 T for IRR0, IRR2.3, IRR3.5, IRR17 and IRR7A. *b)* Maley analysis at µ$_0H$ = 0.5 T for IRR7 and IRR7A. The inset shows the $G(T)$ dependence used for the Maley analysis.

Figure 4. Log-log plot of $J_c$ vs(1$-\left(\dfrac{T}{T_c}\right)^2$) for the samples indicated in the panel at µ$_0H$ = 0.1 T.

**Figure 1**

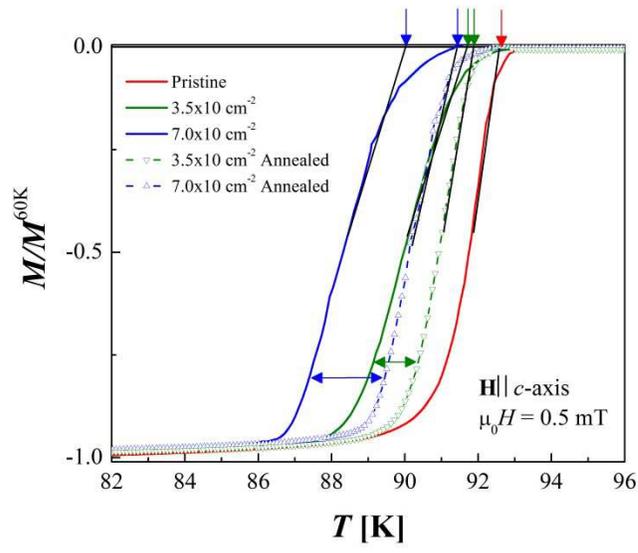

**Figure 2**

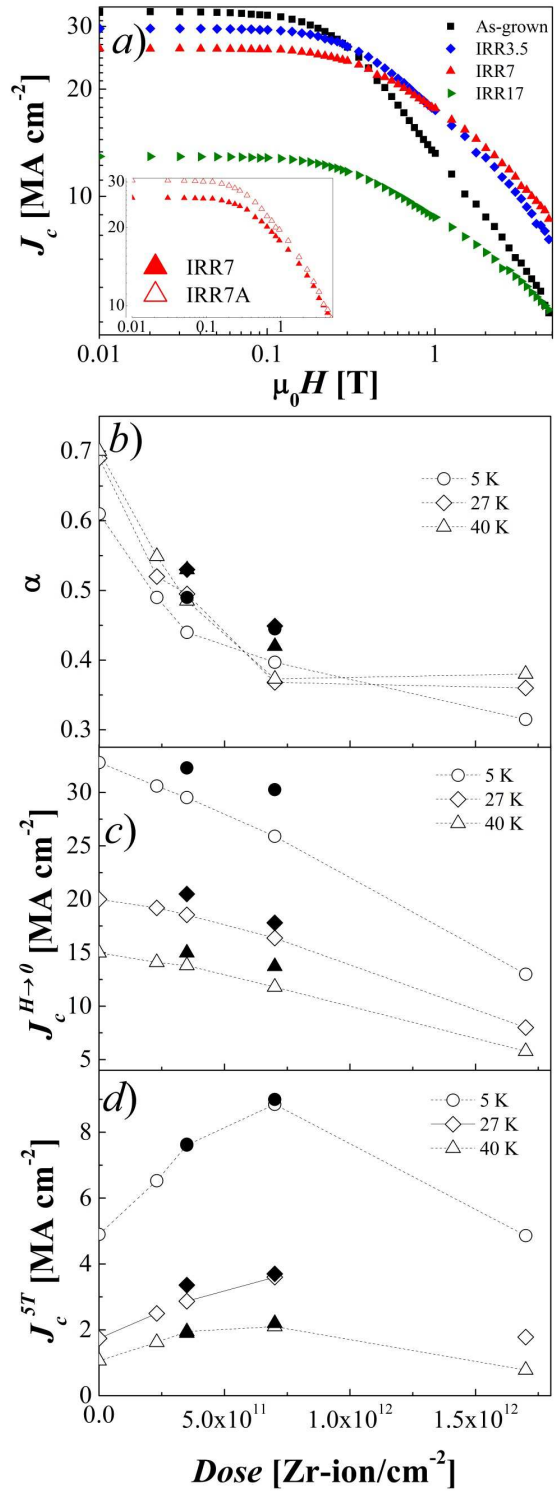

**Figure 3**

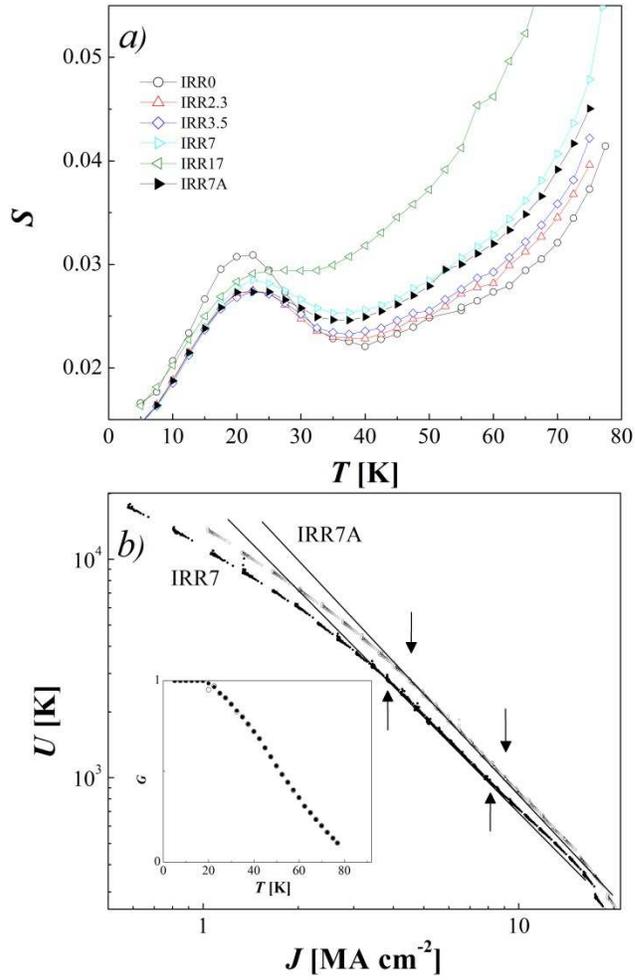

**Figure 4.**

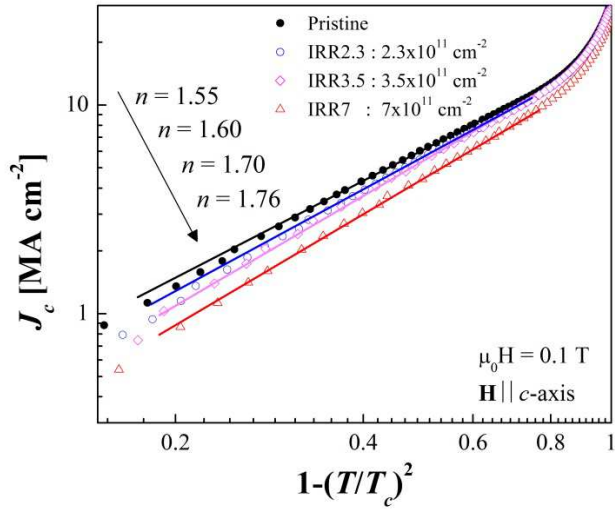

**Table I**

| Sample | $T_c$ [K] | $\mu$ |
|---|---|---|
| Pristine | 92.7 ± 0.2 | 1.63 ± 0.02 |
| IRR2.3 - 2.3x10$^{11}$ cm$^{-2}$ | 92.1± 0.1 | 1.60± 0.01 |
| IRR3.5 - 3.5x10$^{11}$ cm$^{-2}$ | 91.6± 0.1 | 1.52± 0.01 |
| IRR7 - 7.0x10$^{11}$ cm$^{-2}$ | 90.0 ± 0.1 | 1.43 ± 0.01 |
| IRR17 - 1.7x10$^{12}$ cm$^{-2}$ | 87.0 ± 0.5 | 1.28± 0.02 |
| IRR30 - 3x10$^{12}$ cm$^{-2}$ | 83.0 ± 1.0 | - |
| IRR3.5A - 3.5x10$^{11}$ cm$^{-2}$ | 91.7 ± 0.1 | 1.56± 0.01 |
| IRR7A - 7x10$^{11}$ cm$^{-2}$ A | 91.4 ± 0.1 | 1.50 ± 0.01 |